\documentclass[oneside, a4paper, onecolumn, 11pt]{article}
\usepackage[left=2.5cm,top=2.5cm,bottom=2.5cm,right=2.5cm]{geometry}
\usepackage{amsmath}
\usepackage{graphicx}
\usepackage{fancyhdr}
\usepackage{calc}
\usepackage{amssymb}
\usepackage[super,sort&compress,comma]{natbib}
\usepackage{setspace}
\usepackage{amsfonts}
\usepackage{commath}
\usepackage[innercaption]{sidecap}
\usepackage[framemethod=TikZ]{mdframed}
\usepackage{wrapfig}
\usepackage{hyperref}
\usepackage{parskip}
\usepackage{mathrsfs}
\usepackage{lineno}

\newcommand{\Rom}[1]{\expandafter\@slowromancap\romannumeral #1@}
\makeatother
{ \normalfont} 

\newcommand{\m}[3]{#1_{#2 #3}}
\newcommand{\av}[1]{\langle #1 \rangle}
\renewcommand{\Re}{{\rm{Re}}}
\newcommand{\knn}{d_{\rm nn}}

\newcommand{\Ea}{\mathbb{E}(A,P_0,P_1)}
\newcommand{\Ew}{\mathbb{E}(A,G,\Omega)}

\newcommand{\x}{\mathbf x}
\newcommand{\f}{\mathbf f}

\newcommand{\g}{g}
\newcommand{\s}{\mathtt{s}}

\expandafter\def\expandafter\normalsize\expandafter{%
	\normalsize
	\setlength\abovedisplayskip{0pt}
	\setlength\belowdisplayskip{5pt}
	\setlength\abovedisplayshortskip{0pt}
	\setlength\belowdisplayshortskip{5pt}
}
\renewcommand{\baselinestretch}{1.1}
\definecolor{Gray}{gray}{0.75}

\newmdenv[backgroundcolor=Gray, leftmargin = 0pt, rightmargin = 0pt, linewidth = 0pt, roundcorner = 2 pt, innerleftmargin=5pt, innerrightmargin=5pt, innertopmargin=5pt, innerbottommargin=5pt]{Frame}

\begin{document}

\begin{center}
{\color{blue} \LARGE \textbf{Emergent stability in complex network dynamics}}

Chandrakala Meena,$^{1-4,*}$ Chittaranjan Hens,$^{5,6}$ Suman Acharyya,$^{1,2}$ Simcha Haber,$^{1}$ \\[5pt] Stefano Boccaletti$^{7-9}$ \& Baruch Barzel$^{1,2,10,*}$
\end{center}

\begin{enumerate}
\footnotesize{
\item
Department of Mathematics, Bar-Ilan University, Ramat-Gan, Israel 
\item
Gonda Multidisciplinary Brain Research Center, Bar-Ilan University, Ramat-Gan, Israel
\item
Chemical Engineering and Process Development, CSIR-National Chemical Laboratory, Pune, India
\item
Indian Institute of Science Education and Research, Thiruvananthapuram, Maruthamala P. O, Vithura, Kerala 695551
\item
Physics and Applied Mathematics Unit, Indian Statistical Institute, Kolkata, India
\item
Center for Computational Natural Sciences and Bioinformatics, International Institute of Information Technology, Gachibowli, Hyderabad-500032, Telangana, India
\item
CNR - Institute of Complex Systems, Florence, Italy
\item
Moscow Institute of Physics and Technology (National Research University), Moscow, Russian Federation
\item
Universidad Rey Juan Carlos, calle Tulip\'{a}n s/n, 28933 M\'{o}stoles, Madrid, Spain
\item
Network Science Institute, Northeastern University, Boston, MA., USA
}
\end{enumerate}
\begin{itemize}
\footnotesize{
\item[*]
\textbf{Correspondence}.\ C.\ Meena:\ meenachandrakala@gmail.com;\ B.\ Barzel:\ baruchbarzel@gmail.com}
\end{itemize}

\vspace{3mm}
\begin{flushright}
\textit{In memory of Prof.\ Robert May}   
\end{flushright}


\hspace{-3.27mm} 
\textbf{
The stable functionality of networked systems is a hallmark of their natural ability to coordinate between their multiple interacting components. Yet, real world networks often appear random and highly irregular, raising the question of what are the naturally emerging organizing principles of complex system stability. The answer is encoded within the system's stability matrix — the Jacobian — but is hard to retrieve, due to the scale and diversity of the relevant systems, their broad parameter space, and their nonlinear interaction dynamics. Here, we introduce the dynamic Jacobian ensemble, which allows us to systematically investigate the fixed-point dynamics of a range of relevant network-based models. Within this ensemble, we find that complex systems exhibit discrete stability classes. These range from asymptotically unstable, where stability is unattainable, to sensitive, in which stability abides within a bounded range of the system's parameters. Alongside these two classes, we uncover a third asymptotically stable class, in which a sufficiently large and heterogeneous network acquires a guaranteed stability, independent of its microscopic parameters and of external perturbation. Hence, in this ensemble, two of the most ubiquitous characteristics of real-world networks - scale and heterogeneity - emerge as natural organizing principles to ensure fixed-point stability in the face of changing environmental conditions.}
   
The study of complex systems is often directed towards dramatic events, such as cascading failures \cite{Buldyrev2010,Dobson2007,Duan2019,Motter2002,Crucitti2004} or abrupt state transitions. \cite{Achlioptas2009,Gao2016,Boccaletti2016,Boccaletti2006,Danziger2019} In reality, however, these represent the exception rather than the rule. In fact, the truly intriguing phenomenon is that, despite enduring constant perturbations and local obstructions, many systems continue to sustain reliably stable dynamics. \cite{Coyte2015,Sole2001,Brenner2017,Pekora1998} This is achieved in the absence of a detailed design, as indeed, the dynamics of the majority of complex systems are mediated by random, often extremely heterogeneous, networks, comprising a large number of interacting components, and driven by a vast space of microscopic parameters. What then are the roots of this observed stability? 

The answer lies in the system's linear stability matrix, namely its Jacobian $J$, whose principal eigenvalue $\lambda$ determines its response to perturbation. \cite{Arnold1973,Hirsch1974} According to linear stability theory, perturbations may either grow exponentially ($\Re(\lambda) > 0$), capturing instability, or decay exponentially ($\Re(\lambda) < 0$), if the system is stable. The challenge is that the structure of $J$ remains elusive, given the scale, diversity and multiple parameters characterizing real-world complex systems.

To address this we derive the \textit{dynamic Jacobian ensemble}, showing that for a rather broad class of dynamics, stability is determined by a small set of analytically accessible parameters. We further show that this ensemble predicts an emergent stability, asymptoticaly robust in the thermodynamic limit ($N \rightarrow \infty$). Therefore, it offers precisely, the desired natural design principles to ensure complex system stability. \cite{McCann2000,Osullivan2019,Barbier2021}

\vspace{3mm}
{\color{blue} \Large \textbf{Results}}  

\textbf{\color{blue} Fixes-point dynamics}.\
Consider a complex system of $N$ interacting components (nodes), whose dynamic activities $\x(t) = (x_1(t),\dots,x_N(t))^{\top}$ are driven by pairwise interactions, potentially nonlinear. The system's fixed-points $\x_\alpha$ capture static states, which, unperturbed, remain independent of time. The dynamics in the vicinity of these fixed-point can be examined through the system's response to small perturbations $\delta \x(t)$, which, in the linear regime, can be approximated by

\begin{equation}
\dod{\delta \mathbf{x}}{t} = J \delta \mathbf{x} + O(\delta \mathbf{x}^2).
\label{Jacobian}
\end{equation} 

Here $J$, an $N \times N$ matrix, represents the system's Jacobian around $\x_\alpha$, which approximates, through a set of linear equations, the original nonlinear system's dynamics in the perturbative limit, \textit{i.e}.\ small activity changes $\x(t) = \x_{\alpha} + \delta \x(t)$. Hence, $J$'s spectral properties, and specifically its principal eigenvalue $\lambda$, are crucial for characterizing the system's fixed-point behavior. 

Two factors shape $J$ - the system's topology, \textit{i.e}.\ who interacts with whom, and its internal dynamics, namely what is the nature of these interactions:\ 

\textbf{\color{blue} Topology}.\ 
The first ingredient that impacts the structure of $J$ is the network topology $A$, a binary matrix ($\m Aij \in (0,1), \m Aii = 0$), typically sparse and often highly heterogeneous. \cite{Caldarelli2007} Designed to capture the linear response between $i$ and $j$, $J$'s off-diagonal terms vanish if there is no direct $i,j$ link, \textit{i.e}.\ $\m Aij = 0 \Longleftrightarrow \m Jij = 0$ for all $i \ne j$. If, however $\m Aij = 1$, then the relevant term is assigned a \textit{weight} $\m Wij$ that captures the strength of the $i,j$ linear dependence. Together, this leads to

\begin{equation}
J = (A - I) \circ W,
\label{JAIW}
\end{equation}

where the Hadamard product $\circ$ represents matrix multiplication element by element, and $I$ is the $N \times N$ identity matrix. In (\ref{JAIW}) the network structure ($A$) determines the non-vanishing terms in $J$, and $W$ determined their weights. The diagonal entries $\m Jii$ are introduced through the second term, $I \circ W$, where $\m Wii$ quantifies $x_i(t)$'s self-linear dependence.

\textbf{\color{blue} Dynamics - the random matrix paradigm}.\ 
To complete the construction of (\ref{JAIW}) we must assign all weights $W$. In many of the traditional analyses these unknown weights are extracted from two pre-selected probability densities, $P_0(w)$ and $P_1(w)$, for the diagonal and off-diagonal terms, respectively. This gives rise to the Jacobian ensemble $\Ea$, in which one first sets the topology $A$, then extracts weights from $P_0(w)$ and $P_1(w)$; Fig.\ \ref{Illustration}a-c.

As a classic example for this ensemble, we consider May's \cite{May1972} construction, in which $A$ is an Erd\H{o}s-R\'{e}nyi (ER) network, the off-diagonal weights follow $\m Wij \sim \mathcal{N}(0,\sigma^2)$, a zero-mean normal distribution, and the diagonal entries are taken uniformly as $\m Wii = 1$. Hence, the interaction strengths are potentially random, but the self-dynamics are driven by the system's intrinsic relaxation timescales, here normalized to unity. In Methods Section 1 we discuss more detailed constructions, that later built on this random matrix paradigm. 

The $\Ea$ ensemble, described above, has two crucial shortcomings:\ (i) it provides no explicit guidelines on how to connect $P_0(w)$ and $P_1(w)$ with the system's specific nonlinear interactions; (ii) by assigning $A$ and $W$ independently, it ignores the potential interplay between the network structure and $J$'s dynamic weights. This stands in sharp contrast with the frequently observed fact that similar networks potentially exhbit profoundly distinct response patterns. \cite{Barzel2013,Harush2017,Hens2019} How then do we appropriately assign the wights $W$ in (\ref{JAIW}) to capture this interplay between structure and dynamics? 

\vspace{3mm}
{\color{blue} \Large \textbf{The dynamic Jacobian ensemble}}  

To construct predictive $J$ matrices we consider each system's specific interaction mechanisms. For example, in epidemic dynamics, individuals interact through infection and recovery, \cite{PastorSatorras2015,Dodds2003,Brockmann2009} in biological networks, proteins, genes and metabolites are linked through biochemical processes \cite{Karlebach2008,Murray1989,Barzel2011,Barzel2012a} and in population dynamics, species undergo competitive or symbiotic exchanges. \cite{Holling1959,Holland2002,Wodarz2002,Berlow2009} Quite generally, these dynamic mechanisms can be represented by 

\begin{equation}
\dod{x_i}{t} = M_0 \big( x_i(t), \m \f0i \big) + \g \sum_{j = 1}^{N} \m Aij
M_1 \big( x_i(t), \m \f1i \big) \m Gij M_2 \big( x_j(t), \m \f2j \big),
\label{Dynamics}
\end{equation}  

a dynamic framework recently introduced by Barzel and Barab\'{a}si. \cite{Barzel2013} Here $M_0(x_i,\m \f0i)$ captures the self-dynamics of all nodes, and the product function $M_1(x_i,\m \f1i) \times M_2(x_j,\m \f2j)$ describes the $i,j$ pairwise interaction. Each of these functions, $M_q(x, \m \f qi)$, $q = 0,1,2$, is characterized by a set of parameters $\m \f qi$, or - collectively $\f$, capturing rate constants, that may be potentially distributed across the system's components. Hence, the functional form of $M_q(x)$ is uniform throughout the network, yet the specific rates and coefficients $\f$ are node/link specific. In a similar fashion, the global interaction rate $\g$ increases/decreases the strength of \textit{all} interactions, while the specific $i,j$ interaction strength is governed by the potentially diverse weight matrix $\m Gij$. Together, (\ref{Dynamics}) provides a generic template, allowing, by appropriately selecting $M_q(x)$, to cover a range of frequently used models in social, \cite{PastorSatorras2015,Dodds2003} biological \cite{Karlebach2008,Barzel2011,Barzel2012a,Murray1989,Holling1959} and technological \cite{Hayes2004} systems (Fig.\ \ref{TestingGround}; see Methods Section 4 and Supplementary Section 1 for an expanded discussion of Eq.\ (\ref{Dynamics})). 

\textbf{\color{blue} Dynamic Jacobians} (Fig.\ \ref{Illustration}h).\ 
To obtain $J$ we relinquish the random matrix construction $\Ea$, and extract the Jacobian directly from Eq.\ (\ref{Dynamics}). In Supplementary Section 2, we show that this leads to a currently unexplored matrix ensemble, in which the Jacobian weights $W$ in (\ref{JAIW}) are strongly intertwined with the weighted topology $A,G$ via 

\vspace{-3mm}
\begin{equation}
\m Wii \sim C(\f,\g) \knn^{\eta} d_i^{\mu}
\label{Jii}
\end{equation}

for the diagonal weights $\m Jii = -\m Wii$, and

\begin{equation}
\m Wij \sim d_i^{\nu} \m Gij d_j^{\rho}
\label{Jij}
\end{equation} 

for the off-diagonal weights $\m Jij = \m Aij \m Wij$ ($i \ne j$). In (\ref{Jii}) and (\ref{Jij}) $d_i = \sum_{j = 1}^N \m Aij \m Gij$ represents the weighted degree of node $i$, and 

\begin{equation}
\knn = \dfrac{1}{N} \sum_{i = 1}^N \dfrac{1}{d_i} \sum_{j = 1}^N \m Aij \m Gij d_j
\label{knn}
\end{equation}

represents the average weighted degree of a nearest neighbor node. \cite{Gao2016} Together these two parameters, $d_i$ and $\knn$, capture the role of the weighted network topology (Fig.\ \ref{Illustration}g). The four exponents, $\Omega = (\eta,\mu,\nu,\rho)$ are determined by the dynamics, \textit{i.e}.\ the functions $M_q(x)$, hence capturing the role of the system's internal driving mechanisms (Fig.\ \ref{Illustration}e). In case of multiple fixed-points, we have $\Omega_1,\Omega_2$ etc., a potentially distinct exponent set per each fixed-point. The analytical extraction of $\Omega$ from $M_q(x)$ is summarized in Methods Sections 2,3. Finally, the coefficient $C(\f,\g) > 0$ is governed by the rate constants $\f$ and $\g$ in (\ref{Dynamics}), which do not play a role in the scaling exponents $\Omega$ (Fig.\ \ref{Illustration}f).

The resulting dynamic Jacobian in (\ref{Jii}) and (\ref{Jij}), our first key result, is fundamentally distinct from the existing random matrix based constructions. On the one hand, the network structure $A$ continues to determine the non-zero entries, similar to the classic ensemble $\Ea$. Also, the typical magnitude of the diagonal entries depends on the system's rates parameters through $C(\f,\g)$, once again, analogous, albeit not identical, to the selection of $P_0(w)$ in the existing ensemble. However, the similarity ends there, as (\ref{Jii}) and (\ref{Jij}), in contrast to $\Ea$, also capture the role of the system's nonlinearity. Specifically, they predict emergent patterns in the structure of $J$, that are rooted in the interplay between topology and dynamics:\ the degrees $\knn,d_i,d_j$ are extracted from the weighted network topology $A \circ G$ (Fig.\ \ref{Illustration}g), while the scaling exponents $\Omega$ are derived from the dynamic functions $M_q(x)$ (Fig.\ \ref{Illustration}e). 

Therefore, we arrive at a new Jacobian ensemble $\Ew$, which, unlike the random $\Ea$, accounts for the effect of the system-specific nonlinear interaction dynamics. Consequently, in $\Ew$, identical networks may give rise to highly distinctive Jacobian matrices, depending on whether the interactions are, \textit{e.g.}, social, biological or ecological, or even on the specific fixed-point within each type of interaction. This is thanks to the unique set of exponents $\Omega$, characterizing each of these systems/states (Fig.\ \ref{Illustration}h). 

\textbf{\color{blue} Testing $\Ew$}.\ 
To examine predictions (\ref{Jii}) and (\ref{Jij}) we constructed a broad testing ground, including seven relevant dynamic models from different domains:\
Epidemic - the SIS model \cite{PastorSatorras2015,Dodds2003,Brockmann2009} for disease spreading;\
Regulatory - the Michaelis-Menten model \cite{Karlebach2008} for gene regulation;\
Inhibitory - growth suppression in pathogen-host interactions;\ \cite{Wodarz2002}
Biochemical - protein-protein interactions \cite{Barzel2011,Barzel2012a,Murray1989} in sub-cellular networks;\
Population $1,2$ - two models of mutualistic \cite{Holling1959} interactions in population dynamics;\
and finally, Power - load distribution in electric transmission networks. 
Applying each of these dynamics to five different model and relevant empirical networks, we arrive at a total of $35$ combinations of networks/dynamics, upon which we test our predicted $J$-ensemble (a detailed description of all models/networks appears in Supplementary Sections 4 and 7; additional dynamics appear in Supplementary Section 5).

In Fig.\ \ref{TestingGround} we present, for each system, its dynamic equation (blue), and the list of relevant networks upon which it was tested (violet). In some cases the system features several fixed-points, for example, Epidemic (Fig.\ \ref{TestingGround}a) exhibits a healthy state (inactive $\x_0$) and a pandemic state (active $\x_1$). These states are presented using a $3$D visualization. The network is laid out on the $x,y$ plane, and the activities $x_i$ of all nodes are captured by the vertical $z$-axis displacement. Hence under $\x_0$ all nodes remain on the $x,y$ plane ($z = 0$), while in the active state $\x_1$ they all have $x_i > 0$. Finally, we display our predicted dynamic exponents $\Omega = (\eta,\mu,\nu,\rho)$ for each system around its active state (orange); see Supplementary Section 4, where we also derive $\Omega$ for the inactive states.

Perturbing the system around its active fixed-point, we constructed the Jacobian matrix $J$ for each of our $35$ systems (Supplementary Section 7.2). In Fig.\ \ref{JacobianFig} we find that, indeed, the diagonal ($\m Wii$) and off-diagonal ($\m Wij$) weights of our numerically obtained $J$ (blue symbols) follow the predicted scaling of (\ref{Jii}) and (\ref{Jij}) (orange solid lines). For example, in Epidemic we predict $\mu = 1$, while for Regulatory we have $\mu = 0$, both scaling relationships clearly evident in Fig.\ \ref{JacobianFig}b,d. This means that extracting all diagonal terms independently from $P_0(w)$, as in $\Ea$, misses the distinct patterns that arise from the nonlinear Epidemic/Regulatory dynamics. Similarly, the off-diagonal terms are proportional to $d_i^{-1}\m Gij d_j^{0}$ in Epidemic (Fig.\ \ref{JacobianFig}c) and $d_i^{0} \m Gij d_j^{-2}$ in Regulatory (Fig.\ \ref{JacobianFig}e) - once again, in striking agreement with our theoretical predictions (orange solid lines). And yet, in stark contrast with the random construction $\Ea$, where $\m Wij$ are extracted blindly from $P_1(w)$. 

Our analysis further predicts that $\Omega$ depends only on $M_q(x)$, thus independent of network structure $A$, weights $G$, or coefficients $\f$ and $\g$. We examine this in Fig.\ \ref{JacobianFig}, by testing each of our dynamics on a diverse set of networks, with different degree/weight distributions. As predicted, we find that $\eta,\mu,\nu$ and $\rho$ are, indeed, universal, conserved across our diverse model (Erd\H{o}s-R\'{e}nyi, Scale-free 1, Scale-free 2) and relevant empirical (Social 1,2, PPI 1,2, etc.) networks. Hence, $\Omega$ captures the intrinsic, and most crucially, hitherto overlooked, contribution of the nonlinear dynamics to the structure of $J$. 

Together, our derivation demonstrates that:\ (i) Actual $J$ are fundamentally distinct from the commonly used random ensembles; (ii) Contrary to these ensembles, they feature non-random scaling patterns in which topology ($\knn,d_i,d_j$) and dynamics ($\Omega$) are deeply intertwined; (iii) These patterns can be analytically traced to the system's dynamics $M_q(x)$ through Eqs.\ (\ref{Jii}) and (\ref{Jij}), giving rise to our new \textit{dynamic Jacobian ensemble} $\Ew$. Next, we use $\Ew$ to derive the conditions for Eq.\ (\ref{Dynamics})'s dynamic stability. 

\vspace{3mm}
{\color{blue} \Large \textbf{Dynamic stability}}

The dynamic stability around a given fixed-point is governed by $J$'s principal eigenvalue $\lambda$, requiring that $\Re(\lambda) < 0$. To obtain $\lambda$, let us first focus on the role of the network topology $A,G$. The ingredients of $J \in \Ew$, as expressed in Eqs.\ (\ref{Jii}) and (\ref{Jij}), suggest that $\lambda$ is strongly linked to the network's weighted degree density function $P(d)$. This is indicated directly through the dependence on $d_i$ and $d_j$, but also indirectly through the nearest neighbor degree $\knn$, whose magnitude depends on the system's degree-heterogeneity. \cite{Newman2010} For instance, in a randomly wired network we have $\knn = \av{d^2}/\av{d}$, \cite{Yan2017,Gao2016} in which the second moment $\av{d^2}$ increases with $P(d)$'s variance, and consequently with $A,G$'s heterogeneity. In case $P(d)$ is fat-tailed, we have \cite{Newman2010}

\begin{equation}
\knn \sim N^{\beta},
\label{Beta}
\end{equation}

an asymptotic divergence with system size. Hence, $\beta$ helps characterize the network's degree-heterogeneity, being $\beta = 0$ for homogeneous networks, in which $P(d)$ is concentrated around its mean, and $\beta > 0$ for heterogeneous $A,G$, where the variance is unbounded. 

The remaining ingredients in (\ref{Jii}) and (\ref{Jij}) that may impact $\lambda$ are $\Omega = (\eta,\mu,\nu,\rho)$ and $C(\f,\g)$. Combining all three contributions together, we show in Supplementary Section 3 that in $\Ew$, the principal eigenvalue asymptotically follows 

\begin{equation}
\Re (\lambda) \sim N^Q \left(1 - \dfrac{C(\f,\g)}{N^S} \right),
\label{Meena}
\end{equation}  

where $Q = \beta \left(1 + \nu + \rho - \eta/2 \right)$ and

\begin{equation}
S = \beta (\s + \nu + \rho - \mu - \eta).
\label{S}
\end{equation}

In (\ref{S}), the parameter $\s$ depends on the sign of the interactions, being $\s = 1$ under cooperative interactions (positive $\m Jij$), such as in Epidemic or Regulatory, and $\s = 0$ if the interactions are adversarial (negative $\m Jij$), \textit{e.g}, Inhibitory or Biochemical. 

Equations (\ref{Meena})-(\ref{S}), our second key result, uncover the asymptotic behavior of $\lambda$ in the limit of a large complex system $N \rightarrow \infty$. Contrary to $\Ea$, in which $\lambda$ is fully determined by $A$ and $G$, here the exponents $S$ and $Q$ depend also on dynamics, via $\Omega$. Most importantly, these equations have crucial implications regarding the system's fixed-point stability, giving rise to three potential stability classes, uniquely predicted within our dynamic $J$-ensemble (Fig.\ \ref{Illustration}i):

\textbf{Asymptotic instability} ($S > 0$, Fig.\ \ref{Illustration}i, red).\ 
In case $S$ in (\ref{Meena}) is positive, we have, for sufficiently large $N$, $\Re(\lambda) \sim N^{Q} > 0$. Therefore, as the system size $N$ is increased, such states inevitably become unstable.
\textbf{Asymptotic stability} ($S < 0$, Fig.\ \ref{Illustration}i, blue).\
For $S < 0$ we have $N^S \to 0$, the r.h.s.\ of (\ref{Meena}) is dominated by the negative term, and hence $\Re(\lambda) < 0$. Consequently, here as $N \rightarrow \infty$ stability becomes unconditionally guaranteed.   
\textbf{Sensitive stability} ($S = 0$, Fig.\ \ref{Illustration}i, green).\
Under $S = 0$ the system lacks an asymptotic behavior, and therefore, its stability depends on $C(\f,\g)$ in (\ref{Meena}). If $C(\f,\g) > 1$ the system is stable, otherwise it becomes unstable. Hence, in this class stability is not driven by the system size $N$, but rather by the coefficient $C(\f,\g)$, and consequently by Eq.\ (\ref{Dynamics})'s rate parameters $\f$ and $\g$.   

\textbf{\color{blue} Stability classifier}.\ 
The \textit{stability classifier} $S$ in (\ref{S}) helps group all $J \in \Ew$ into distinct stability classes. It achieves this by identifying the relevant topological ($\beta$) and dynamic ($\Omega$) control parameters that help analytically predict the stability of any system within the form of Eq.\ (\ref{Dynamics}). We can therefore use $S$ to predict \textit{a priori} whether a specific combination of topology and dynamics will exhibit stable functionality or not.   

To examine $S$'s predictive power, we tested it extensively against a diverse set of complex networks. Specifically, we used our model and empirical networks to extract $7,387$ Jacobian matrices from the $\Ew$ ensemble, with different sets of $\eta,\mu,\nu$ and $\rho$. In Fig.\ \ref{Classes}a we show the principal eigenvalue $\Re(\lambda)$ vs.\ $S$ for the entire $7,387$ Jacobian sample. As predicted, we find that the parameter $S$ sharply splits the sample into three classes. The asymptotically unstable class (red, top-right) has $S > 0$ and consequently also $\Re(\lambda) > 0$, a guaranteed instability. The asymptotically stable class (blue, bottom-left) is observed for $S < 0$, and has, in all cases $\Re(\lambda) < 0$, \textit{i.e}.\ stable dynamics. Finally, for $S = 0$ we observe sensitive stability, with $\Re(\lambda)$ having no asymptotic behavior, positive or negative (green). A small fraction ($\sim 8\%$) of our sampled $J$ matrices were inaccurately classified by $S$ (grey), an expected consequence of the approximate nature of $S$'s derivation (Supplementary Section 3).

\vspace{3mm}
{\color{blue} \Large \textbf{The ingredients of dynamic stability}}

The parameter $S$ in (\ref{S}) reduces Eq.\ (\ref{Dynamics})'s dynamic stability into five relevant exponents. The first four $\Omega = (\eta,\mu,\nu,\rho)$ are determined by the system's intrinsic dynamics $M_q(x)$, around each of its fixed-points. The remaining exponent in (\ref{S}), $\beta$, is independent of the dynamics, determined solely by $A,G$, specifically by their weighted degree density function $P(d)$, through (\ref{Beta}). Therefore, together, $S$ captures the roles of both topology and dynamics, whose interplay determines the system's stability class around a specific fixed-point - stable, unstable or sensitive.

The only remaining factor in (\ref{Meena}) is the coefficient $C(\f,\g)$, whose value is driven by Eq.\ (\ref{Dynamics})'s rate parameters $\f,\g$. Yet, as our analysis indicates, this factor is sidelined when $N \to \infty$ under $S \ne 0$. We interpret this to mean that under asymptotic stability or instability, the system's countless microscopic parameters turn irrelevant, and the stable/unstable fixed-points of (\ref{Dynamics}) become ingrained into the system's intrinsic dynamics, \textit{i.e}.\ the functional form of $M_q(x)$; see Methods Section 4 and Supplementary Section 1 for an expanded discussion on this distinction.  

To gain deeper insight, consider, for example, the factors that drive a system towards the loss of stability. Most often such events result from external stress or changes in environmental conditions. \cite{Gao2016} Such forces impact the system by perturbing its dynamic parameters, \textit{e.g}., changing the rates of specific processes. Seldom, however, do these environmental perturbations affect the system's built-in interaction mechanisms. Indeed, these mechanisms are ingrained in the \textit{physics} of the interacting components, and therefore they are unaffected by external conditions. Hence, asymptotic stability ($S < 0, N \to \infty$) depicts robust dynamic states, that are insensitive to changes in environmental conditions.  

\textbf{\color{blue} The role of degree-heterogeneity}.\
The dependence of $S$ in (\ref{S}) on $\beta$ highlights the crucial role that $P(d)$ plays in dynamic stability. To understand this, consider a homogeneous network, such as ER with randomly assigned weights. Here $P(d)$ follows a Poisson distribution, having $\beta = 0$. Under these conditions we have $S = 0$ in (\ref{S}), the system has no defined asymptotic behavior, and hence it is sensitively stable - \textit{i.e}.\ its stability depends on model parameters via $C(\f,\g)$. Hence, our predicted asymptotically stable/unstable classes depend on $\beta > 0$, indicating that they emerge as a direct consequence of degree-heterogeneity. This suggests that a fat-tailed $P(d)$, indeed - among the defining features of many real-world complex systems,\cite{Caldarelli2007} serves as a dynamically stabilizing structure, locking-in specific fixed-points, in the face of a persistently fluctuating environment.

To further uncover the roots of asymptotic stability/instability, we consider again $\Ew$'s principal eigenvalue $\lambda$ in (\ref{Meena}). Its structure portrays stability as a balance between the positive, \textit{i.e}.\ destabilizing, effect mediated by the network interactions, vs.\ the negative, stabilizing, feedback, driven by the parameter $C(\f,\g)$ in $J$'s diagonal in Eq.\ (\ref{Jii}); Fig.\ \ref{Classes}b. It is, therefore, natural to enhance stability by increasing $C(\f,\g)$, which, in effect, translates to strengthening each node's intrinsic self regulation. Equation (\ref{Meena}) predicts that $J$ becomes stable if $C(\f,\g)$ exceeds a critical value

\begin{equation}
C_0 \sim N^S,
\label{C0}
\end{equation} 

beyond which $\lambda$ turns negative. For asymptotically stable states ($S < 0$) we have, for sufficiently large $N$, $C_0 \rightarrow 0$, a guaranteed stability even under arbitrarily small $C(\f,\g)$. In contrast, for asymptotically unstable states ($S > 0$) we have $C_0 \rightarrow \infty$, hence such systems are impossible to stabilize even under extremely large $C(\f,\g)$. We emphasize that $C(\f,\g)$ is the only component in (\ref{Meena}) that is dependent on the system's tunable parameters, and therefore having an unbounded range of $C$-values under which the system remains stable (or unstable) guarantees that $\lambda$ is, indeed, unaffected by parameter perturbation, \textit{e.g}., changing environmental conditions.

To test Eq.\ (\ref{C0}), in Fig.\ \ref{Classes}c-k we extract a set of three specific $J$ matrices from $\Ew$, representing systems from our three stability classes:\ $J_{\rm AS}$, asymptotically stable with $\Omega = (2,2,2,-1)$; $J_{\rm SS}$, sensitively stable with $\Omega = (0,-1,-2,0)$; and $J_{\rm AU}$, asymptotically unstable with $\Omega = (1,-2,-1,2)$. For each of these we plot $C_0$ vs.\ $N$, capturing the level of negative feedback required to ensure the system's stability. Under ER ($\beta = 0$, Fig.\ \ref{Classes}i-k) we do not observe a defined asymptotic behavior. The critical $C_0$ does not scale with $N$, indicating that sufficient perturbation to the model parameters can, indeed, affect $J$'s stability. 

In contrast, the same $J$ matrices on our scale-free network SF1 ($\beta = 0.6$) exhibit a clear asymptotic behavior, congruent with prediction (\ref{C0}). For $J_{\rm AS}$ we have $C_0 \sim N^{-1.2}$, while under $J_{\rm AU}$ we observe $C_0 \sim N^{1.8}$ (Fig.\ \ref{Classes}f-h, squares), precisely as predicted (solid lines). Finally, in $J_{\rm SS}$, having $S = 0$, the system, again lacks an asymptotic behavior, and therefore can be stabilized (or destabilized) under finite $C_0$, independently of system size $N$ (Fig.\ \ref{Classes}g). 

Together, Eq.\ (\ref{C0}) helps us link the scale $N$ of a complex system with its observed stability. As opposed to the random matrix viewpoint of $\Ea$, in which $N$ has a destabilizing effect, and hence large systems become unstable, \cite{May1972} our dynamic ensemble uncovers broad conditions where the contrary is true, and $N \to \infty$, is, in fact, what anchors the system's stability (Fig.\ \ref{FigMay}). Next, we return to our testing ground of dynamical systems (Fig.\ \ref{TestingGround}) to examine this asymptotic stability, not just on artificially constructed $J$, but under the full nonlinear setting of Eq.\ (\ref{Dynamics}). 

\textbf{\color{blue} Emergent stability}.\
The stabilizing/destabilizing effect of $N$ and $P(d)$ is especially relevant if (\ref{Dynamics}) exhibits multiple fixed-points, for example, an undesirable $\x_0$ and a desirable $\x_1$. In $\Ew$ these two states can be potentially characterized by two different exponent sets $\Omega_0$ and $\Omega_1$, and consequently a different stability profile. If \textit{e.g}., $\x_0$ is asymptotically unstable ($S > 0$) and $\x_1$ is asymptotically stable ($S < 0$), then a large ($N \to \infty$) heterogeneous ($P(d)$ fat-tailed) network will firmly reside only in $\x_1$, unaffected by perturbation to $\f$ or $\g$.

To observe this we return to our testing ground of Fig.\ \ref{TestingGround}, this time focusing on dynamic models that have multiple fixed-points. This includes Regulatory, Epidemic and Inhibitory, each of which exhibits on top of its \textit{active} state $\x_1$, in which all $x_i > 0$, an \textit{inactive} state $\x_0 = (0,\dots,0)^\top$, where all activities vanish (Population 1,2 also exhibit an inactive $\x_0$, however it is never stable, see Supplementary Section 4.3). 

First, we simulated Regulatory on an ER network, and varied the model's two parameters $f_i$, the individual node degradation rate, and $\g$, the global interaction strength. We find that when the average $\av f$ is large or, alternatively, when $\g$ is small the system resides in $\x_0$, whereas in the opposite limit, it favors $\x_1$ (Fig.\ \ref{Emergent}c-e, diamonds). This is precisely the \textit{sensitive stability}, in which the system's fixed-point behavior is driven by its microscopic parameters. Repeating the same experiment on our scale-free network SF1, we observe that $\x_1$ is sustained for a broader range of $\av f$ and $\g$, hence SF1 is comparably insensitive to changes in these parameters (Fig.\ \ref{Emergent}d-f, triangles). This robustness is a direct outcome of our classifier:\ $\x_0$ has $\Omega_0 = (0,0,0,0)$, which in ({\ref{S}) predicts $S = \beta > 0$, while $\x_1$ has $\Omega_1 = (0,0,0,-2)$, and hence $S = -\beta < 0$. Therefore, on a large ($N \to \infty$) scale-free ($\beta > 0$) network, $\x_0$ becomes asymptotically unstable, and the system is forced to reside in the asymptotically stable $\x_1$.   

To observe this systematically we seek the critical global weight $\g_c$, below which $\x_1$ becomes unstable, and the system transitions to $\x_0$. Varying the system size $N$ over $4$ orders of magnitude, from $10$ to $2 \times 10^4$, we observe first hand $\x_1$'s asymptotic stability:\ while under ER $\g_c$ is almost independent of $N$ (Fig.\ \ref{Emergent}g, diamonds), in SF1 it scales negatively with system size, approaching $\g_c \rightarrow 0$ in the limit $N \rightarrow \infty$ (triangles). Hence, as predicted, SF1's $\x_1$ state remains stable even under arbitrarily small $\g_c$, a stability entrenched by system size. This reconfirms prediction (\ref{C0}), but this time, not on theoretically constructed $J$ from $\Ew$, as shown in Fig.\ \ref{Classes}f-k, but rather on the actual numerically simulated dynamics of Eq.\ (\ref{Dynamics}). Similar stability patterns are also observed in Epidemic and Inhibitory (Fig.\ \ref{Emergent}h,i and Supplementary Sections 4.1 and 4.5).

\textbf{\color{blue} The role of the hub nodes}.\ 
We, therefore, observe a qualitative difference between homogeneous vs.\ fat-tailed $P(d)$, in which degree-heterogeneity can potentially afford the network a guaranteed stability, that is asymptotically independent of microscopic parameters. This phenomenon is rooted in the dominance of the hub nodes, whose dynamic behavior forces the entire system towards stability/instability. In that sense, one can think of our classifier $S$ as a mathematical tool to predict precisely what will be the dynamic role of the hubs - whether the hubs serve as stabilizers ($S < 0$), destabilizers ($S > 0$) or neither ($S = 0$).

\vspace{3mm}
{\color{blue} \Large \textbf{Discussion and outlook}}

The linear stability matrix $J$ carries crucial information on the dynamic behavior of complex systems. Here, we exposed distinct patterns in the structure of $J$ that arise from the nature of the system's interaction dynamics. These patterns are expressed through the four dynamic exponents $\Omega = (\eta,\mu,\nu,\rho)$, which we link analytically to the system's dynamic functions, $M_q(x)$, independently of the weighted network topology $A,G$ or parameters $\f,\g$. We interpret this to mean that $\Omega$ is hardwired into the system's innate interaction dynamics, determined by the dynamic \textit{model}, \textit{e.g}., Epidemic or Regulatory, but not by the specific model parameters or the system's underlying connectivity patterns. Therefore, our predicted Jacobian ensemble in (\ref{Jii}) and (\ref{Jij}), as well as its associated stability classifier $S$ in (\ref{S}), both capture highly robust and distinctive characteristics of the system's dynamics, that cannot be perturbed or otherwise affected by shifting environmental conditions. 

Graph spectral analysis represents a central mathematical tool to translate network structure into dynamic predictions. \cite{Almendral2007,VanMieghem2012,VanMieghem2010} A network's spectrum, \textit{i.e}.\ its set of eigenvalues and eigenvectors, captures information on its dynamic timescales, potential states, and - in the present context - its dynamic stability. Most often, spectral analysis is applied to the network \textit{topology}, namely we seek the \textit{graph's} eigenvalues, thus overlooking information on the nonlinear dynamics that occur \textit{on} that graph. As an alternative, our $\Ew$ ensemble suggests to apply spectral analysis, not to the topology $A$ (or the weighted $A \circ G$), but rather to $J$, which, thanks to $\Omega$ preserves the information of both structure and dynamics.

Strictly speaking, our analysis covers the Barzel-Barab\'{a}si family of equations in (\ref{Dynamics}). With that said, it also shows strong numerical indications for broader relevance beyond this family (Methods Section 5; Supplementary Section 5). Most importantly, it motivates a departure from the decades old random matrix paradigm ($\Ea$), by showing that real-world Jacobians are anything but random. Hence, while the analytically predictable scaling patterns observed here are specific to Eq.\ (\ref{Dynamics}), the notion that such patterns dominate the structure of $J$ is, likely, much more general, and should be pursued as a systematic road map by which to analyze complex system dynamics. 

\vspace{3mm}

{\color{blue} \textbf{Acknowledgments}}.\
We wish to thank Itamar Conforti for designing inspiring artwork to accompany our scientific research. C.M.\ thanks the Planning and Budgeting Committee (PBC) of the Council for Higher Education, Israel for support. C.M. is also supported by the INSPIRE-Faculty grant (code:\ IFA19-PH248) of the Dept.\ of Science and Technology, India.
C.H.\ is supported by the INSPIRE-Faculty grant (code:\ IFA17-PH193) of the Dept.\ of Science and Technology, India. S.H. has contributed to this work while visiting the Mathematics Department of Rutgers University, New Brunswick.
S.B.\ acknowledges funding from the project EXPLICS granted by the Italian Ministry of Foreign Affairs and International Cooperation. This research was also supported by the Israel Science Foundation (grant No.\ 499/19), the Israel-China ISF-NSFC joint research program (grant No.\ 3552/21), the US National Science Foundation-CRISP award No.\ 1735505, and by the Bar-Ilan University Data Science Institute grant for data science research.

{\color{blue} \textbf{Author contribution}}.\
All authors designed and planned the research and derived its analytical results. CM, with the aid of CH and SA, conducted the data analysis and numerical simulations. BB was the lead writer of the paper. 

{\color{blue} \textbf{Competing interests}}.\
The authors declare no competing interests.

\clearpage

\begin{figure}[ht]
\caption{
\textbf{The dynamic Jacobian ensemble}.\ 
To predict dynamic stability we seek the system's stability matrix $J$. 
(a)-(b) The classic approach is to structure $J$ around the network topology $A$, with weights extracted from two distributions:\ $P_0(w)$ for the diagonal entries $\m Jii$ and $P_1(w)$ for the interactions strengths $\m Jij$, $i \ne j$.   
(c) This provides $J \in \Ea$, a random-matrix based construction, whose stability is determined by the structure $A$ and the random weights $W$.\
(d) The dynamic $J$ ensemble features emergent patterns that arise from:\
(e) The functional form of $M_q(x)$ (orange), capturing the system's ingrained dynamics, \textit{e.g}., social, biological or technological.\ We derive $\Omega$ in Eqs.\ (\ref{Jii}) and (\ref{Jij}) directly from these three functions. 
(f) The microscopic parameters $\f,\g$ (turquoise) that provide the specific rate-constants for (\ref{Dynamics})'s dynamic processes. For example, the infection rate in Epidemic (top), or the degradation rate in Biochemical (bottom). These parameters are tunable, following changes in social behavior (Epidemic) or temperature (Biochemical). Their impact on $J$ is encapsulated within the coefficient $C(\f,\g)$ in (\ref{Jii}).
(g) $A,G$ represent the weighted network (purple), expressed in $J$ via the density function $P(d)$, the nearest-neighbor degree $\knn$ in (\ref{knn}) and $\beta$ in (\ref{Beta}).
(h) The resulting $J$-ensemble, $\Ew$, exhibits non-random scaling patterns. Similar to the random $\Ea$, the non-vanishing terms correspond to the network links, however, in contrast to the random weights of $\Ea$, here the weight of the $i,j$ entry depends on $d_i$ and $d_j$, as well as on $\Omega$ (orange captions). The result is a dynamic ensemble, in which identical networks ($A,G$) yield highly distinctive $J$ matrices depending on $M_q(x)$, \textit{e.g.}, Regulatory (left), Epidemic (center) or Biochemical (right).    
(i) The stability of $J$ boils down to the classifier $S$ in (\ref{S}), whose value depends on degree heterogeneity ($\beta$, purple) and on $\Omega$ (orange terms), but not on the parameters $\f,\g$. Therefore, it provides a robust classification into stable (blue) or unstable (red) dynamics, asymptotically insensitive to changes in $\f,\g$. Under $S = 0$ the system becomes sensitive (green), and stability is driven precisely by $\f,\g$.
}
\label{Illustration}
\end{figure}
 
\clearpage

\begin{figure}[ht]
\caption{
\textbf{Testing ground for the $\Ew$ ensemble}.\ 
We constructed different combinations of (weighted) networks $A,G$ and dynamics $M_q(x)$ to examine our $J$-ensemble. 
(a) Epidemic.\ We implemented the susceptible-infected-susceptible (SIS) dynamics (grey box) on a set of model and real-world networks (violet box). This system exhibits two potential fixed-points (3D plots):\ inactive $\x_0$, in which all activities vanish, \textit{i.e}.\ healthy, and active $\x_1$, where all $x_i > 0$, namely pandemic. In this 3D visualization the nodes $i = 1,\dots,N$ are laid out on the $x,y$ plane and their fixed-point activity $x_i$ is represented by color (red - low, blue - high) and vertical displacement ($z$-axis). Therefore, in $\x_0$ all nodes are on the $x,y$ plane ($z = 0$), and in $\x_1$ they are distributed along $z > 0$ and range from red to blue. In each of these states the system has a different set of exponents $\Omega$ and hence a different Jacobian $J$. Here we present $\Omega$ and $S$ for the non-vanishing state $\x_1$ (orange box). The remaining panels follow a similar format. 
(b) Regulatory.\ Sub-cellular dynamics following the Michaelis-Menten model. Here $\Omega,S$ depend on the model exponents $a,h$.
(c) Population 1.\ Mutualistic interactions in, \textit{e.g}., microbial communities.
(d) Inhibitory.\ Suppression dynamics, \textit{e.g.}, between hosts and pathogens.
(e) Biochemical.\ Protein-protein interactions modeled via mass-action kinetics. This system exhibits a single fixed-point.
(f) Power.\ Synchronization dynamics between power system components.
(g) Population 2. Mutualistic population dynamics with non-additive interactions, namely replacing the term $\sum_{j = 1}^N \m Aij \m Gij M_2(x_j)$ in Eq.\ (\ref{Dynamics}) by $M_2(\sum_{j = 1}^N \m Aij \m Gij x_j)$. 
(h) Our networks, including Erd\H{o}s-R\'{enyi} and scale-free with both normally and power-law distributed weights, together with relevant empirical networks. A detailed description of all networks appears in Supplementary Section 7.4. Together we arrive at a set of $35$ combinations of networks/dynamics upon which we test our theoretical framework. A detailed analysis of all dynamic models appears in Supplementary Section 4. Note that Population 2 and Power (panels f,g) are not in the form of Eq.\ (\ref{Dynamics}), and hence they expand our testing ground beyond the bounds of our analytical framework (Supplementary Section 5).
}
\label{TestingGround}
\end{figure} 
 
\clearpage

\begin{figure}[ht]
\caption{
\textbf{Emergent patterns in the dynamic ensemble $\Ew$}.\ 
We implemented our seven dynamic models, Epidemic, Regulatory etc., on relevant model and empirical networks, as detailed in Fig.\ \ref{TestingGround}; see legend at bottom. Perturbing all nodes around their numerically obtained fixed-point ($\x_1$) we constructed the Jacobian $J$. 
(a) The numerical simulations incorporate the full complexity of Eq.\ (\ref{Dynamics}):\ weighted network (purple), diverse parameters (turquoise) and nonlinear mechanisms (orange). This provides actual Jacobian matrices, obtained from numerical runs of the nonlinear network models (Simulation, blue, top). We compare our simulation results with our predictions in (\ref{Jii}) and (\ref{Jij}) (Theory, orange, bottom).
(b) The diagonal weights $\m Wii$ vs.\ $d_i$ as obtained from Epidemic dynamics (symbols). We observe our predicted scaling (\ref{Jii}) with $\mu = 1$ (orange solid line). The scaling is independent of $A,G$, observed consistently on all our model/empirical networks - intrinsic to the Epidemic dynamics, as predicted.
(c) The off-diagonal weights $\m Wij$ vs.\ our theoretical prediction of (\ref{Jij}) with $\nu = -1, \rho = 0$ (symbols). Once again, we observe a perfect agreement between simulation (blue symbols) and theory (orange solid line). We also include two relevant empirical networks, Social 1 and Social 2 (light blue circles/squares), capturing online social dynamics. 
(d)-(e) Similar results are observed under Regulatory ($\mu = \nu = 0, \rho = -2$) on both model and empirical networks (PPI 1 and PPI 2);
(f)-(g) Population 1 dynamics ($\mu = \nu = 1, \rho = -2$, empirical networks:\ Microbial 1 and Microbial 2);
(h)-(i) Inhibitory dynamics ($\mu = 1, \nu = 1/2, \rho = -1$, empirical networks:\ PPI 1 and PPI 2);
(j)-(k) Biochemical dynamics ($\mu = 1, \nu = -1, \rho = 0$, empirical networks:\ PPI 1 and PPI 2).
(l)-(m) Power dynamics ($\mu = 1, \nu = \rho = 0$, empirical networks:\ Power 1 and Power 2);
(n)-(o) Population 2 dynamics ($\mu = 0, \nu = -2, \rho = 0$, empirical networks:\ Microbial 1 and Microbial 2).
In all systems, we find that real Jacobian matrices (blue symbols) are well-approximated by our theoretically predicted scaling laws (orange solid lines). Data in all panels are logarithmically binned. \citep{Milojevic2010} Details on numerical calculation of $J$, log-binning and all networks appear in Supplementary Section 7.   
}
\label{JacobianFig}
\end{figure} 

\clearpage

\begin{figure}[ht]
\caption{
\textbf{Three classes of dynamic stability}.\ 
We extracted $7,387$ Jacobian matrices from the $\Ew$ ensemble, using combinations of model/empirical networks with different dynamic exponents $\Omega$. For each $J$ we calculated the principal eigenvalue $\lambda$ and the stability classifier $S$ in (\ref{S}).
(a) $\Re(\lambda)$ vs.\ $S$ for all $7,387$ $J$-matrices. We observe the three predicted classes:\ Asymptotically unstable (red) in which $S > 0$ and hence, as predicted, we have also $\Re(\lambda) > 0$; Sensitively stable (green), where $S = 0$ and $\Re(\lambda$ can be both positive or negative; Asymptotically stable (blue), where $S < 0$ and therefore $\Re(\lambda) < 0$. Our classification showed $\sim 4\%$ inaccuracy on binary networks, $\sim 5\%$ on weighted networks and $\sim 15\%$ on weighted/negative networks - a total discrepancy of $\sim 8\%$ (grey dots) over the entire ensemble.  
(b) The value of $\lambda$ emerges from the competition between $J$'s off-diagonal terms, representing positive feedback, and the strength of the diagonal terms $\m Jii$ (negative feedback). Therefore one can force a system towards stability ($\Re(\lambda) < 0$) by increasing the coefficient $C$ in (\ref{Jii}).  
(c)-(e) Taking three specific $J$ matrices, we plot $\Re(\lambda)$ vs.\ $C$, seeking the critical $C_0$, in which $\Re(\lambda)$ becomes negative (grey lines). This represents the critical $C$, above which stability is ensured.
(f) For the stable $J_{\rm AS}$ ($S = -1.2$) we find that $C_0$ decreases with $N$ (squares), capturing the asymptotic stability, in which as $N \rightarrow \infty$ stability is sustained even under arbitrarily small $C$. The theoretical scaling predicted in Eq.\ (\ref{C0}) is also shown (solid line, slope $-1.2$). 
(g) For $J_{\rm SS}$ we have $S = 0$, the critical $C_0$ is independent of $N$, hence the system's stability can be affected by finite changes to its dynamic parameters.
(h) The asymptotically unstable $J_{\rm AU}$ ($S = 1.8$) has $C_0 \rightarrow \infty$ in the limit of large $N$, in perfect agreement with Eq.\ (\ref{C0}) (solid line). Here, no matter how large is $C$, the fixed-point associated with $J_{\rm AU}$ is always unstable.
(i)-(k) For a homogeneous $P(d)$, \textit{e.g}., Erd\H{o}s-R\'{e}nyi, $\beta$ vanishes and hence $S = 0$ in (\ref{S}). Under these conditions, regardless of $\Omega$, the system is always sensitively stable and therefore $C_0$ does not scale with $N$. This demonstrates the role of degree-heterogeneity for ensuring stability in the face of changing environmental conditions.  
}
\label{Classes}
\end{figure} 

\clearpage

\begin{figure}[ht]
\caption{
\textbf{Will a large complex system be stable?} 
This question, first posed by May in 1972, \cite{May1972} captures a long standing challenge, fueled by the seeming contradiction between theory and practice. 
(a) While empirical reality answers with an astounding yes, May's mathematical analysis, based on random matrix theory, suggested the contrary, that large systems are inevitably unstable, giving rise to the well-known \textit{diversity-stability debate}. Here, the series of growing networks (left to right) becomes increasingly unstable as we drift towards $N \to \infty$. In the works that followed it became clear that real-world complex systems are not random. Rather they incorporate unique structural \cite{Allesina2012,Tarnowski2020,Sinha2005} and dynamic \cite{Kirk2015,Osullivan2019,Barbier2021} constraints - or organizing principles - that can potentially enhance stability. 
(b) Our dynamic Jacobian ensemble offers such organizing principles, that emerge quite naturally in a variety of real-world systems (Fig.\ \ref{TestingGround}). This is expressed through the built-in scaling patterns in $J$ (orange), which, in turn, predict a broad class of asymptotically stable dynamic states (middle). In this class ($S < 0$) system size plays a stabilizing, rather than a destabilizing role. Consequently, we arrive at quite broad conditions where May's original question receives a clear answer:\ large complex systems not only can, but, often must be stable.} 
\label{FigMay}
\end{figure}

\clearpage

\begin{figure}[ht]
\caption{
\textbf{Emergent stability in large heterogeneous networks}.\
(a)-(b) Regulatory dynamics exhibit two fixed points (3D plots), each with its own $J \in \Ew$, shown to the right of each plot. The inactive $\x_0$ has $\Omega = (0,0,0,0)$ and $S = \beta > 0$ - hence it is asymptotically unstable. The active $\x_1$ has a different $J$, with $\Omega = (0,0,0,-2)$ and $S = -\beta < 0$, asymptotically stable.
(c) The state of Regulatory as obtained from numerical simulations on our Erd\H{o}s-R\'{e}nyi (ER) network under varying $\g$. The system transitions from $\x_1$ (right) to $\x_0$ (left) under small $\g$. This represents sensitive stability, as indeed predicted for ER, in which parameters, here $\g$, affect the state of the system.
(d) To examine this systematically we plot the mean activity $\av \x$ vs.\ $\g$ as obtained for ER (diamonds) and for our scale-free network SF1 (triangles). Both systems exhibit a critical $\g_c$, below which $\x_1$ becomes unstable and the system transitions to $\x_0$. The crucial point is, however, that thanks to its heterogeneity SF1 exhibits an increased robustness against $\g$ variations, with $\g_c$ an order of magnitude lower than that observed for ER.
(e) $\av \x$ vs.\ $\av f$ shows a similar behavior (here increasing $\av f$ causes the system to collapse to $\x_0$). 
(f) The state of SF1 under the same four conditions shown in panel (c). As predicted, SF1 remains at $\x_1$ even when ER has already collapsed to $\x_0$.
(g) $\g_c$ vs.\ the system size $N$ as obtained from numerically simulating Regulatory dynamics. For ER we observe $\g_c \sim \rm const$ (diamonds), hence there is a typical $\g_c$ below which the system transitions to $\x_0$. Consequently $\x_1$ can be destabilized via parameter perturbation. Note that while $N$ spans over four orders of magnitude, $\g_c$ varies by a mere $\sim 40\%$. In contrast, for SF1 we find that $\g_c$ exhibits negative scaling with $N$, approaching $\g_c \to 0$ in the limit $N \to \infty$. This captures precisely the predicted asymptotic stability, in which a sufficiently large and heterogeneous network is guaranteed to stably reside in $\x_1$ even under arbitrarily small $\g$ (or large $\av f$).
(h)-(i) Repeating this experiment for Epidemic and Inhibitory, we continue to observe our predicted asymptotic stability:\ under SF1 we have $\g_c \to 0$ as $N \to \infty$, whereas under ER $\g_c$ is (almost) independent of $N$.     
}
\label{Emergent}
\end{figure}

\nolinenumbers

\clearpage
{\color{blue} \Large \textbf{References}}
\vspace{3mm}
\renewcommand{\baselinestretch}{0.5}
\renewcommand{\section}[2]{}
\bibliographystyle{unsrt}
\begin{footnotesize}

\end{footnotesize}
 
\clearpage

{\color{blue} \Large \textbf{Methods}}

{\color{blue} \textbf{1.\ Random matrix based Jacobian constructions}}

The random matrix paradigm was first introduced by May, \cite{May1972} seeking precisely the question we address here (Fig.\ \ref{FigMay}):\ will a large complex system be stable? In this original construction all diagonal weights  in (\ref{JAIW}) were set to $\m Wii = 1$, while the off-diagonal weights were extracted from a zero-mean Gaussian distribution. The rationale is that the self regulation of all components is uniform, driven by the system's intrinsic timescales (normalized to unity), while the interaction strengths vary randomly around zero. Such construction is a particular case of our $\m Wii$ in (\ref{Jii}), setting $C(\f,g) = 1$ and $\eta = \mu = 0$. While the first assumption about $C(\f,g)$ has no significant bearing on our analysis, the second, which ignores the dynamic exponents $\eta,\mu$, is precisely the crux of our proposed novelty. Indeed, in our framework, it is these two exponents (together with $\nu$ and $\rho$) that capture the role of the nonlinear dynamics, ignored in the random matrix constructions.   

In the works that followed May's abstract construction, researchers systematically introduced more realism into $J$. First, by considering more realistic $\m Aij$, for example, small-world \cite{Watts1998} or scale-free networks, \cite{Yan2017} which have, indeed, been shown to impact (\ref{JAIW})'s spectral properties. Other advances tackled $P_0(w)$ and $P_1(w)$, showing that different dynamics may lead to more specific weight distributions, rather than the originally assumed Gaussian distribution. This is achieved by conditioning $P_0(w)$ and $P_1(w)$ to account for specific patterns that arise from known dynamic processes. For example, in predator prey relationships a positive $\m Wij$ is often matched with a negative $\m Wji$, \cite{Allesina2012} capturing the asymmetry in the benefit/loss of the predator and its prey. More complex dynamic constraints may further impact the statistical properties of $W$, limiting the Jacobian to a selected subset of the random matrix ensemble. \cite{Kirk2015}

{\color{blue} \textbf{2.\ Deriving the dynamic Jacobian ensemble}}

While we provide a complete and rigorous derivation of the $\Ew$ ensemble in Supplementary Sections 1-3, below we include a shorthand version of this derivation, tracking the main steps and important mathematical transitions leading to Eqs.\ (\ref{Jii}), (\ref{Jij}) and (\ref{S}). For simplicity, in this abbreviated analysis, we limit ourselves to systems with uniform weights/parameters. Hence, in Eq.\ (\ref{Dynamics}) we set the global and individual weights to $\g = \m Gij = 1$, and take $\m \f qi = \m \f qj$ for all $i,j = 1,\dots,N$. Under these simplifications, we rewrite Eq.\ (\ref{Dynamics}) as

\begin{equation}
\dod{x_i}{t} = M_0 \big( x_i(t) \big) + M_1 \big( x_i(t) \big) \sum_{j = 1}^{N} \m Aij M_2 \big( x_j(t) \big),
\label{Dynamics2}
\end{equation}   

omitting the link weights $\g,\m Gij$ and the parameters $\m \f qi$, which are now identical for all nodes. We emphasize that in our full derivation, as well as in our reported results and simulations, we do not rely on these simplified assumptions, and only employ them here for brevity and conciseness. 

{\color{blue} \textbf{Fixed-point analysis}}.\
Starting from (\ref{Dynamics2}), we seek the system's potential fixed-points via

\begin{equation}
M_0 (x_i) + M_1 (x_i) \sum_{j = 1}^{N} \m Aij M_2 (x_j) = 0,
\label{FixedPoint}
\end{equation} 
 
where we use $x_i$ (omitting the $t$ dependence) to denote the fixed-point $x_i = x_i(t \to \infty)$. To express the summation over $j$ in the l.h.s.\ we use 

\begin{equation}
\av{M_2(x)}_{i\odot} = 
\dfrac{1}{d_i} \sum_{j = 1}^{N} \m Aij M_2(x_j),
\label{M2odoti}
\end{equation}

capturing a weighted average over $M_2(x_j)$ across all of $i$'s nearest neighbors. Here, with all link weights set to unity, $d_i = \sum_{j = 1}^N \m Aij$ represents $i$'s binary degree. This generalizes to $i$'s weighted degree if we reintroduce our weights $\g,\m Gij$. In (\ref{M2odoti}) we use the notation $\odot$ to represent a \textit{neighborhood average}, namely an average over $i$'s surrounding nodes $i\odot$. Substituting (\ref{M2odoti}) into (\ref{FixedPoint}) we obtain

\begin{equation}
M_0 (x_i) + d_i M_1 (x_i) \av{M_2(x)}_{i\odot} = 0,
\label{FixedPoint2}
\end{equation} 

which we further simplify to 

\begin{equation}
R(x_i) = q_i
\label{Rxiqi}
\end{equation}

where $R(x) = - M_1(x)/M_0(x)$ and

\begin{equation}
q_i = \dfrac{1}{\av{M_2(x)}_{i\odot} d_i}
\label{qi}
\end{equation}

is node $i$'s inverse weighted degree. In Eq.\ (\ref{Rxiqi}), the function $R(x)$ is only defined in case $M_0(x) \ne 0$. The treatment of $M_0(x) = 0$ is done separately in Supplementary Section 2.5. We can now extract the fixed-point $x_i$ by inverting $R(x_i)$ to obtain

\begin{equation}
x_i = R^{-1}(q_i),
\label{xiRInv}
\end{equation} 

allowing us to express the fixed-point activity of node $i$ in function of its inverse degree $q_i$. Here, we rely on the implicit assumption that $R(x)$ is invertible, allowing us to write $R^{-1}(q_i)$ in (\ref{xiRInv}). As above, we employ this assumption here only for simplicity; in our complete derivation in Supplementary Section 2, we show how to obtain $x_i$ also under non-invertible $R(x)$.

{\color{blue} \textbf{Jacobian scaling - diagonal weights $\m Wii$}}.\ 
We now return to Eq.\ (\ref{Dynamics2}) to extract the Jacobian weights $\m Wii$ and $\m Wij$ around the fixed-point obtained in (\ref{xiRInv}). Starting with the diagonal terms, we write 

\begin{equation}
\m Wii = \left. 
\dfrac{\partial \dot{x}_i}{\partial x_i} 
\right|_{x_i = R^{-1}(q_i)} =
\left. \left(
M_0^\prime(x_i) + M_1^\prime(x_i) \sum_{j = 1}^N \m Aij M_2(x_j)
\right) \right|_{x_i = R^{-1}(q_i)},
\label{Wii}
\end{equation}

where $M_q^\prime(x) = \partial M_q / \partial x$. Equation (\ref{Wii}) represents a derivative of the r.h.s.\ of (\ref{Dynamics2}) taken around the fixed-point $x_i$, which we express via (\ref{xiRInv}) as $R^{-1}(q_i)$. Next we use $R(x) = -M_1(x)/M_0(x)$ to write $M_0(x) = -M_1(x)/R(x)$, allowing us to express the first derivative on the r.h.s.\ of (\ref{Wii}) as

\begin{equation}
M_0^\prime(x_i) = - \dfrac{M_1^\prime(x_i)}{R(x_i)} + \dfrac{M_1(x_i)R^\prime(x_i)}{R^2(x_i)},
\label{M0primex}
\end{equation}

which, setting $x_i = R^{-1}(q_i)$, provides

\begin{equation}
M_0^\prime(x) = - \dfrac{M_1^\prime \big( R^{-1}(q_i) \big)}{q_i} +
\dfrac{M_1 \big( R^{-1}(q_i) \big) 
R^\prime \big( R^{-1}(q_i) \big)}{q_i^2}.
\label{M0primex2}
\end{equation}

To obtain the denominators, $q_i$ and $q_i^2$ on the r.h.s.\ of (\ref{M0primex2}) we used the fact that $R(R^{-1}(q_i)) = q_i$. We can now use (\ref{M2odoti}) to express the sum on the r.h.s.\ of (\ref{Wii}) as $\sum_{j = 1}^N \m Aij M_2(x_j) = d_i \av{M_2(x)}_{i\odot}$, which, according to (\ref{qi}) is equal to $1/q_i$. Collecting all the terms we arrive at

\begin{equation}
\m Wii = - \dfrac{M_1^\prime \big( R^{-1}(q_i) \big)}{q_i} +
\dfrac{M_1 \big( R^{-1}(q_i) \big) 
R^\prime \big( R^{-1}(q_i) \big)}{q_i^2} + 
\dfrac{M_1^\prime \big( R^{-1}(q_i) \big)}{q_i},
\label{Wii2}
\end{equation}

which in turn provides

\begin{equation}
\m Wii = \dfrac{1}{q_i^2} Y \big( R^{-1}(q_i) \big),
\label{Wii2}
\end{equation}

where $Y(x) = M_1(x)R^\prime(x)$. 

Equation (\ref{Wii2}) expresses the diagonal Jacobian weight $\m Wii$ in terms of $i$'s inverse degree $q_i \sim d_i^{-1}$. In the asymptotic limit of large $d_i$ (small $q_i$) we can approximate (\ref{Wii2}) by expanding $Y(R^{-1}(q_i))$ around $q_i = 0$. We, therefore, express this function as a Hahn \cite{Hahn1995} power series expansion in the form

\begin{equation}
Y \big( R^{-1}(q_i) \big) = \sum_{n = 0}^{\infty} B_n q_i^{\m \Phi_n},
\label{YRinv}
\end{equation}

allowing us below to examine the limit $q_i \to 0$. The Hahn series in (\ref{YRinv}) represents a generalization of the Taylor expansion to allow for negative and real powers, hence $\Phi_n \in \mathbb{R}$ captures a sequence of real powers in ascending order, \textit{i.e}.\ $\Phi_0 < \Phi_1$ and so on. In the limit $q_i \to 0$ we take only the leading term $q_i^{\Phi_0}$, which in (\ref{Wii2}) provides the scaling relationship

\begin{equation}
\m Wii \approx B_0 q_i^{-\mu} = 
B_0 \Big( \av{M_2(x)}_{i\odot} d_i \Big)^{\mu},
\label{Wii3}
\end{equation}

where $\mu = 2 - \Phi_0$. In the last step of (\ref{Wii3}) we reintroduced $d_i$ using the definition of $q_i$ in (\ref{qi}), hence also adding the $i$ neighborhood average $\av{M_2(x)}_{i\odot}$.

Equation (\ref{Wii3}) describes the weight of the diagonal Jacobian entry associated with a specific node $i$. It is found to depend on the node's degree $d_i$, but also on the activity of its neighboring nodes via $\av{M_2(x)}_{i\odot}$. To complete the scaling of $\m Wii$ with $i$'s degree $d_i$ we must characterize the $d_i$-dependence of $\av{M_2(x)}_{i\odot}$. The crucial point is that $\av{M_2(x)}_{i\odot}$ captures an average over $i$'s \textit{neighborhood}, not over the node $i$ itself, and hence, on average, it is only indirectly affected by $i$'s degree $d_i$. To express this more rigorously we write 

\begin{equation}
\av{M_2(x)}_{i\odot} \approx \av{M_2(x)}_{\odot} f(d_i),
\label{EnsAve}
\end{equation}

replacing the average over $i$'s neighborhood ($i\odot$) with the \textit{ensemble} average ($\odot$). This ensemble average $\av{M_2(x)}_{\odot} = (1/N) \sum_{i = 1}^N \av{M_2(x)}_{i\odot}$ represents an aggregation over \textit{all} nodes in the network, and hence it is independent of $i$ or $d_i$. To account for the potential $d_i$ dependence, we include, on the r.h.s.\ of (\ref{EnsAve}), the implicit function $f(d_i)$. This function, defined as $f(d_i) = \av{M_2(x)}_{i\odot} / \av{M_2(x)}_{\odot}$, captures the distinction between the conditional $i$-neighborhood average ($i\odot$) vs.\ the network's ensemble average ($\odot$). It, therefore, helps quantify potential statistical dependencies between $i$ and its interacting neighbors $i\odot$. Hence, if the network is randomly wired, \textit{i.e}.\ lacks degree-correlations, \cite{Newman2010} we have $f(d_i) = 1$, independently of $i$. However, if correlations are present, it will be expressed through a non-trivial $f(d_i)$.

Extracting only the terms that depend on $d_i$ we rewrite Eq.\ (\ref{Wii3}) as $\m Wii \sim f^\mu (d_i) d_i^\mu$, omitting the terms $B_0,\av{M_2(x)}_{\odot}$, which are independent of $d_i$. Finally, if $f(d_i)$ is sub-polynomial, it does not contribute to the $d_i$ scaling in the limit of large $d_i$. This allows us to write

\begin{equation}
\m Wii \sim d_i^\mu,
\label{Wiidimu}
\end{equation} 

recovering the asymptotic scaling relationship of Eq.\ (\ref{Jii}). In (\ref{Wiidimu}) we eliminated all terms that do not contribute to the polynomial dependence on $d_i$, thus focusing solely on the obtained \textit{scaling} relationship. These terms may, however, depend on other parameters of (\ref{Dynamics}). For example, quite expectedly the term $\av{M_2(x)}_{\odot}$, an average driven by the activity of all nearest neighbor nodes, is, potentially dependent on the nearest neighbor degree $\knn$ in (\ref{knn}). Similarly, the coefficient $B_0$ is, most often, a function of the parameters $\f$ and $\g$ in (\ref{Dynamics}). These additional dependencies are precisely what gives rise the pre-factors $C(\f,\g) \knn^\eta$ in (\ref{Jii}), which we ignored in the present derivation (see Supplementary Section 2 for the complete derivation, which covers also these terms). 

The substitution leading to Eqs.\ (\ref{EnsAve}) and (\ref{Wiidimu}) represents our first approximation, where we assume that $\av{M_2(x)}_{i\odot}$ is only \textit{weakly} dependent on $i$'s degree $d_i$. This weak dependence is precisely defined by the assumption that $f(d_i)$ is sub-polynomial, \textit{e.g}., $f(d_i) \sim \log d_i$. This implies that the neighbors of a node $i$ with degree $d_i$ are, to a sufficient degree, statistically similar to those of $j$ whose degree is $d_j$. Under this approximation, averaging over a node's neighborhood, conditional on that node's degree, as we do in the r.h.s.\ of (\ref{Wii3}) is (almost) the same as averaging over the neighbors of any other node, independently of degree (indeed, up to the sub-polynomial correction $f(d_i)$). In Supplementary Section 1.2 we elaborate on the relevance of this approximation, and in Supplementary Fig.\ 2 we explicitly measure $f(d_i)$ for our entire testing ground of networks/dynamics. We find that $f(d_i$) is, indeed, at most logarithmic, supporting the relevance of our approximation for our set of real/model networks.  

{\color{blue} \textbf{Off-diagonal weights $\m Wij$}}.\
To extract the off-diagonal terms $i \ne j$ of $\Ew$ we return to Eq.\ (\ref{Dynamics2}), this time writing

\begin{equation}
\m Wij = \left. 
\dfrac{\partial \dot{x}_i}{\partial x_j} 
\right|_{\substack{x_i = R^{-1}(q_i) \\ x_j = R^{-1}(q_j)}} =
\left. 
\dfrac{\partial}{\partial x_j} 
\left(
M_0(x_i) + M_1(x_i) \sum_{n = 1}^N \m Ain M_2(x_n)
\right) \right|_{\substack{x_i = R^{-1}(q_i) \\ x_j = R^{-1}(q_j)}}.
\label{Wij1}
\end{equation}

Keeping only the terms that explicitly depend on $x_j$ we obtain

\begin{equation}
\m Wij = 
M_1(x_i) \m Aij M_2^\prime(x_j) 
\Big|_{\substack{x_i = R^{-1}(q_i) \\ x_j = R^{-1}(q_j)}} = 
M_1 \big( R^{-1}(q_i) \big) \m Aij M_2^\prime \big( R^{-1}(q_j) \big), 
\label{Wij2}
\end{equation}

helping us identify the two relevant dynamic functions $M_1 \big( R^{-1}(q_i) \big)$ and $M_2^\prime \big( R^{-1}(q_j) \big)$, whose leading powers determine the scaling of $\m Wij$. Expressing these functions as Hahn series we write

\begin{eqnarray}
M_1 \big( R^{-1}(q_i) \big) &=& \sum_{n = 0}^{\infty} K_n q_i^{\Pi_n}
\label{M1Hahn}
\\
M_2^\prime \big( R^{-1}(q_j) \big) &=& \sum_{n = 0}^{\infty} L_n q_j^{\Theta_n},
\label{M2primeHahn}
\end{eqnarray} 

and in the limit of large $d_i$ and $d_j$ (small $q_i,q_j$) take only the leading terms $\sim q_i^{\Pi_0}$ and $\sim q_j^{\Theta_0}$. Substituting these terms into (\ref{Wij2}), and using the fact that $q_i \sim d_i^{-1}$, we arrive at

\begin{equation}
\m Wij \sim d_i^\nu \m Aij d_j^\rho,
\label{Wijdidj}
\end{equation} 

where $\nu = - \Pi_0$ and $\rho = - \Theta_0$, recovering the prediction of Eq.\ (\ref{Jij}), under the current setting of $\m Gij = 1$, \textit{i.e}.\ unweighted.

The obtained exponents $\mu,\nu$ and $\rho$ are all extracted from the leading powers of our derived dynamic functions $Y(R^{-1}(x))$ in (\ref{YRinv}), $M_1( R^{-1}(x))$ in (\ref{M1Hahn}) and $M_2^\prime ( R^{-1}(x))$ in (\ref{M2primeHahn}). These functions, in turn, are directly linked to $M_q(x)$ in (\ref{Dynamics}), and hence offer a direct procedure by which to extract the $\Ew$ Jacobian scaling relationships, as outlined in Fig.\ \ref{Illustration}. The forth and final exponent $\eta$ in (\ref{Jii}) can be extracted in a similar fashion, as we show in Supplementary Section 2. 

{\color{blue} \textbf{3.\ Practical summary - calculating $\Omega$}}

While the derivation in Methods Section 2 may be elaborate, its practical outcome is rather straightforward, providing a step-by-step recipe by which to construct the exponent set $\Omega = (\eta,\mu,\nu,\rho)$ in Eqs.\ (\ref{Jii}) and (\ref{Jij}). First, we use the dynamic functions $M_0(x),M_1(x)$ and $M_2(x)$ of Eq.\ (\ref{Dynamics}) to construct the three secondary functions 

\begin{equation}
\begin{array}{ccc}
R(x) = -\dfrac{M_1(x)}{M_0(x)}, & 
Y(x) = M_1(x) R^{\prime}(x), & 
Z(x) = R(x)M_2(x).
\end{array}
\label{FunctionsFrame}
\end{equation}

The functions $R(x)$ and $Y(x)$ are introduced in Methods Section 1 above; $Z(x)$ is derived in Supplementary Section 2. From (\ref{FunctionsFrame}) we extract four additional functions, which we express through a Hahn power-series expansion as

\begin{equation}
\begin{array}{cc}
M_2 \big( Z^{-1}(x) \big) = \displaystyle 
\sum_{n = 0}^{\infty} G_n x^{\Psi_n}, & 
Y \big( R^{-1}(x) \big) = \displaystyle
\sum_{n = 0}^{\infty} C_n x^{\Phi_n}, 
\\ \\
M_1 \big( R^{-1}(x) \big) = \displaystyle 
\sum_{n = 0}^{\infty} K_n x^{\Pi_n}, &
M_2^{\prime} \big( R^{-1}(x) \big) = \displaystyle 
\sum_{n = 0}^{\infty} L_n x^{\Theta_n}
\end{array}.
\label{HahnFrame}
\end{equation}

We use $R^{-1}(x)$ and $Z^{-1}(x)$ to denote the inverse functions of $R(x)$ and $Z(x)$. The leading powers ($n = 0$) in these Hahn series directly provide $\Omega$ via

\begin{equation}
\begin{array}{cccc}
\mu = 2 - \Phi_0, & 
\nu = -\Pi_0, &
\rho = -\Theta_0, &
\eta  = -\Psi_0 (\mu - \nu - \rho).
\end{array}
\label{ExponentsFrame}
\end{equation}

Hence, to construct $J \in \Ew$ we first generate the weighted network $A \circ G$, then extract the weighted degrees $d_i,d_j$ of all nodes and the nearest neighbor degree $\knn$ of Eq.\ (\ref{knn}). The resulting $J$ satisfies 
\begin{eqnarray}
\m Jii &\sim& -C(\f,\g) \knn^{\eta} d_i^{\mu}
\label{JiiFrame}
\\[8pt]
\m Jij &\sim& d_i^{\nu} \m Aij \m Gij d_j^{\rho},
\label{JijFrame}
\end{eqnarray}

where the coefficient $C(\f,\g) > 0$ encapsulates the system's specific rate parameters (we do not attempt to predict this coefficient in the current formalism). The detailed derivation of $\Omega$ appears in Supplementary Section 2, followed by a step by step application on all our testing ground dynamics (Fig.\ \ref{TestingGround}) in Supplementary Section 4. 

In the above formulation we have assumed that $R(x)$ and $Z(x)$ are invertible, writing $R^{-1}(x)$ and $Z^{-1}(x)$ in (\ref{HahnFrame}). In Supplementary Section 2 we explain how to properly treat non-invertible $R(x),Z(x)$. In these sections, we also demonstrate how to extract $J$ for system's with multiple fixed-points, and, specifically, in Supplementary Section 2.5, how to construct $J$ around a trivial fixed-point $\x = (0,\dots,0)^{\top}$.

{\color{blue} \textbf{4.\ The ingredients of $\Ew$}}

In Eq.\ (\ref{Dynamics}) we distinguish between the nonlinear form of the functions $M_q(x)$ and their specific parameters $\m \f qi$. The former, we argue, are designed to mathematically represent the nodes' intrinsic driving mechanisms, distinguishing between, \textit{e.g}., Epidemic vs.\ Biochemical dynamics. The latter, on the other hand, describes the specific rates of these mechanistic processes, which may, potentially change across nodes/links, or under different environmental conditions. To root this distinction on mathematical grounds we refer again to the Hahn expansion, and express each of the functions $M_q(x, \m \f qi)$ via
\begin{equation}
M_q(x_i, \m \f qi) = \sum_{n = 0}^{\infty} \m Cqn(\m \f qi) x_i^{\m \Gamma qn}.
\label{Hahn}
\end{equation} 

In (\ref{Hahn}) we distinguish between the role of the \textit{powers} $\m \Gamma qn$ and that of the \textit{coefficients} $\m Cqn$. The powers, in most cases, characterize the \textit{functional form} of $M_q(x, \m \f qi)$, differentiating, for example, between $M_q(x, \m \f qi) \sim x^2$ or $M_q(x, \m \f qi) \sim x/(1 + x)$. These different functions are designed to represent, mathematically, distinct microscopic mechanisms, \textit{e.g}., social interactions vs.\ biological processes. As these mechanisms are ingrained into the \textit{physics} of the interacting components, we take them, in our formulation, to be fixed and uniform across all nodes/links. In contrast, the coefficients $\m Cqn$ are often tunable, depending on the particular rates characterizing each node's dynamics, and hence they depend on the node specific parameters $\m \f qi$. 

To better understand this distinction let us consider a specific example of logistic growth, a common mechanism in population dynamics. Within the dynamic framework of Eq.\ (\ref{Dynamics}) this mechanism is captured by $M_0(x_i) = b_i x_i (1 - x_i/c_i)$, which written in the form (\ref{Hahn}), provides $M_0(x_i) = b_i x_i - (b_i/c_i)x_i^2$. Namely the coefficients are $\m C00 = b_i$ and $\m C01 = -b_i/c_i$, and the corresponding powers are $\m \Gamma 00 = 1$ and $\m \Gamma 01 = 2$. The crucial point is that while $b_i$ and $c_i$, \textit{i.e}.\ the species growth rate and the system's carrying capacity, are node dependent, and potentially affected by environmental conditions, the functional form $M_q(x_i) \sim x_i(1 - x_i)$ is intrinsic to logistic growth, and cannot be easily perturbed. This is precisely captured by the separate role of powers vs.\ coefficients:\ the tunable parameters $b_i,c_i$ are expressed only within $\m Cqn$, whereas the logistic growth functional form is embedded within the powers $\m \Gamma qn$ - here describing linear growth ($\m \Gamma 00 = 1$) followed by quadratic attenuation due to intra-species competition ($\m \Gamma 01 = 2$). 

Hence, from a strictly mathematical perspective, we define \textit{parameters} as the factors affecting the coefficients $\m Cqn$ in (\ref{Hahn}), and \textit{functional form} via the set of participating powers $\m \Gamma qn$. Our interpretation of this mathematical distinction is that the powers are more intrinsic than the coefficients. Indeed, in our logistic growth example, the two powers arise from the system's ingrained driving mechanisms - of growth (linear) vs.\ competition (quadratic). In contrast, the coefficients depend on the parameters $b_i,c_i$, which may assume any value within the logistic growth framework, and can even change due to external conditions. 

The crucial point is that our Jacobian scaling exponents $\Omega$ depend only on the powers $\m \Gamma qn$, and are unrelated to the coefficients $\m Cqn$. Hence, in our example, all systems driven by logistic growth (and a matching interaction dynamics) will have similar $\Omega$, regardless of the specific parameters $b_i,c_i$. This portrays $\Omega$ and its resulting $S$, as an innate built-in characteristic of the system's dynamics, detached from its multitude of microscopic parameters. Consequently, our asymptotic stable/unstable classes are intrinsic to the system's dynamics, insensitive to external perturbation or to microscopic discrepancies.

{\color{blue} \textbf{5.\ Generality and limitations the $\Ew$ ensemble}}

\textbf{\color{blue} Dynamic limitations}.\
Our ensemble $\Ew$ was analytically derived under the conditions defined by the Barzel-Barab\'{a}si equation (\ref{Dynamics}). Despite its general structure, we wish to emphasize that this equation still excludes several families of dynamics. For example, non-additive interactions or threshold models. \cite{Granovetter1978} Similarly, if the system incorporates a mixture of distinct interaction mechanisms, such that every node/link is driven by its own idiosyncratic processes, the dynamics cannot be cast into the form $M_0(x),M_1(x),M_2(x)$. 

We note, however, that while our analytical derivations are, indeed, bounded by these restrictions, the family of potential dynamics included within the $\Ew$ ensemble may, in fact, be broader. Specifically, in Supplementary Section 5 we consider several expansions to Eq.\ (\ref{Dynamics}) that help us examine the applicability limits of our dynamic Jacobians:\

\begin{itemize}
\item
\textbf{Non-factorizable interactions}.\
Our testing ground includes Power dynamics, in which the interaction term cannot be partitioned into a product $M_1(x_i)M_2(x_j)$, but rather incorporates a diffusive mechanism of the form $M(x_j - x_i)$. Such dynamics, excluded from (\ref{Dynamics}), arise in different contexts, from reaction-diffusion to synchronization, \cite{Kuramoto1984,Kundu2018} and despite the fact that they are not covered by our anaytical framework, our analysis of Power indicates that they continue to fall within $\Ew$.

\item
\textbf{Non-additive interactions}.\
Another outlier in Fig.\ \ref{TestingGround} is Population 2, in which the linear sum $\sum_{j = 1}^N M_2(x_j)$ is replaced by $M_2(\sum_{j = 1}^N x_j)$, again - outside the bounds of (\ref{Dynamics}). Still, as shown, \textit{e.g}., in Fig.\ \ref{JacobianFig}n,o, this system also has $J \in \Ew$. 

\item
\textbf{Mixed dynamics}.\
The last assumption we challenge is the notion that all components are driven by similar dynamic processes, as expressed by the uniform functional form of $M_q(x)$ across all nodes. In Supplementary Sections 5.4 and 5.5 we examine, numerically, systems with two or three competing self or interaction dynamics. The Jacobians of such systems, we find, exhibit coexisting scaling relationships with exponent sets $\Omega_1,\Omega_2,\dots$, corresponding to the network's distinct dynamic mechanisms. This captures a natural generalization of $\Ew$, that indicates the potential qualitative insight offered by our analysis, even beyond Eq.\ (\ref{Dynamics})'s technical limits.
\end{itemize}

\textbf{\color{blue} Topological limitations}.\
Our predicted asymptotic stability/instability is driven by the limit of large $d$, \textit{i.e}. the hubs. It is therefore mainly relevant for degree-heterogeneous networks. While extreme heterogeneity is, indeed, common in many biological and social systems, there are areas, such as in ecological systems, \cite{Stouffer2005} were the networks tend to be more homogeneous. Under such conditions, our theory predicts that the system is in the sensitive class:\ it \textit{could} be stable, but its stability is not guaranteed in the face of parameter perturbation.   

Finally, our asymptotic predictions capture the system's global stability, but have no bearing on the dynamic stability of small motifs or sub-networks, which may be locally unstable. Still in an asymptotically stable system, the global impact of such unstable motifs, vanishes in the limit of large $N$, and hence the system as a whole remains insensitive to these local discrepancies. We discuss this in detail, including extensive numerical support in Supplementary Section 6.     

\vspace{3mm}

{\color{blue} \textbf{Data availability}}.\
All empirical network data to retrieve the results shown here is available online at:\ 
\href{https://gitlab.com/meenachandrakala/Dynamic_Stability/-/tree/master/Dynamic_Stability}{https://gitlab.com/meenachandrakala/Dynamic\_Stability/-/tree/master/Dynamic\_Stability}.

{\color{blue} \textbf{Code availability}}.\
All code to reproduce the results shown here is available online at:\ 
\href{https://gitlab.com/meenachandrakala/Dynamic_Stability/-/tree/master/Dynamic_Stability}{https://gitlab. \\ com/meenachandrakala/Dynamic\_Stability/-/tree/master/Dynamic\_Stability}.

\end{document}